\documentclass[11pt,a4paper]{article}

\usepackage{amsmath}
\usepackage{amsthm}
\usepackage{amsfonts}
\usepackage{graphicx}
\usepackage{enumerate}
\usepackage{fullpage}
\usepackage{cite}
\usepackage{hyperref}

\newcommand{\pn}[1]{} 

\newcommand{\beq}{\begin{equation}}
\newcommand{\eeq}{\end{equation}}
\newcommand{\beqn}{\begin{eqnarray}}
\newcommand{\eeqn}{\end{eqnarray}}

\newcommand{\nn}{\nonumber}

\newcommand{\bt}[1]{| #1 \rangle}
\newcommand{\Z}[1]{\mathbb{Z}_{#1}}

\newcommand{\sixt}{$\textbf{16}$}

\newcommand{\sixtbar}{$\overline{\textbf{16}}$}

\def\AEF{A.E. Faraggi}
\def\JHEP#1#2#3{{\it JHEP} {#1} (#2) #3}

\def\NPB#1#2#3{{\it Nucl.\ Phys.}\/ {B #1} (#2) #3}
\def\PLB#1#2#3{{\it Phys.\ Lett.}\/ {B #1} (#2) #3}

\def\PRT#1#2#3{{\it Phys.\ Rep.}\/ {\bf#1} (#2) #3}

\newcommand{\lefto}{left}
\newcommand{\righto}{right}

\renewcommand{\L}{L}
\newcommand{\R}{R}
\newcommand{\twozero}{$(2,0)$}

\def\heteroticconvention{0}  

\if0\heteroticconvention
\fi

\if1\heteroticconvention
\renewcommand{\lefto}{right}
\renewcommand{\righto}{left}

\renewcommand{\L}{R}
\renewcommand{\R}{L}
\newcommand{\twozero}{$(0,2)$}
\fi

\begin{document}
\begin{titlepage}
\thispagestyle{plain}
\pagestyle{plain}      
\rightline{LTH-1000} 
\vspace{1.5cm}

\begin{center}
 {\Large \bf 
Spectral flow as a map between N=(2,0)-models}
\end{center}

\begin{center}

P. Athanasopoulos$^*$\footnote{
		                  E-mail address: panos@liv.ac.uk}, 
A.E. Faraggi$^*$\footnote{
		                  E-mail address: alon.faraggi@liv.ac.uk} and
D. Gepner$^\dagger$\footnote{
                                  E-mail address: doron.gepner@weizmann.ac.il}
\\

\bigskip

{$^*$} { Department of Mathematical Sciences,\\
		University of Liverpool,     \\
                Liverpool L69 7ZL, United Kingdom}\\

\smallskip
                
{$^\dagger$} { Department of Particle Physics, \\ 
         Weizmann Institute, Rehovot 76100, Israel\\}  
\end{center}

\smallskip

\begin{abstract}
The space of (2,0) models is of particular interest among all 
heterotic-string models because it includes the models
with the minimal $SO(10)$
unification structure, which is well motivated by the
Standard Model of particle physics data.
The fermionic $\Z{2}\times \Z{2}$ heterotic-string models revealed 
the existence of a new symmetry in the space of string
configurations under the exchange of spinors and vectors of the 
$SO(10)$ GUT group, dubbed spinor-vector duality. In this paper we generalize this idea to arbitrary internal 
rational Conformal Field Theories (RCFTs). We explain how
the spectral flow operator normally acting within a general $(2,2)$
theory can be used as a map between $(2,0)$ models. We describe the details,
give an example and propose more simple currents that can be used in 
a similar way.
\end{abstract}

\end{titlepage}
\setcounter{page}{2}

\section{Introduction}\label{sec:intro}

String theory provides a detailed framework to explore the unification of
the gauge and gravitational interactions \cite{books}.
The construction of phenomenological
models that can make contact with the real world has been of great interest
and understanding their underlying structure can be especially elucidating. 
However, the vastness of the {\it a priori} possible vacuum solutions
impedes progress towards the construction of a standard string model. 
In this respect, the study of various relationships between 
different models can be very fruitful. In particular, it may have
far reaching implications for the interpretation of the landscape
of string vacua. 

In the heterotic constructions the left and right moving sectors are 
treated asymmetrically. Of particular interest are the so called \twozero{}
models\footnote{Our convention here is that the \lefto-moving sector is 
supersymmetric and the \righto-moving is bosonic. }
because it is known that $N=1$ spacetime supersymmetry requires
(at least) \twozero{} world-sheet supersymmetry and because
they can accommodate $SO(10)$ unification. The problem is that the
space of these models is huge. For example, even though the number of
$(2,2)$ Gepner models\cite{Gepner} is quite tractable and 
they have been studied in detail\cite{Font1990a, Schellekens1990}, 
the number of \twozero{} models that arise is much
greater\cite{Schellekens1990}. 
For this reason it would be very useful 
to discover relations in the space of such models.

In this paper we will make a small step in this direction by getting 
inspiration from a new kind of duality that comes under the name spinor-vector 
duality and was observed in $\Z{2}\times\Z{2}'$ orbifold 
models\cite{Faraggi2007a,Faraggi2007b,Faraggi2008,Catelin-Jullien2009,
Angelantonj2010,Faraggi2011}.
It is a duality of the massless spectra of two such models
under the exchange of vectorial and spinorial representations of the $SO(10)$
GUT gauge group. 

These models turn out to be related through the spectral flow operator and
the underlying CFT structure of the spinor-vector duality for
$\Z{2}\times\Z{2}'$ orbifolds was discussed in \cite{Faraggi2011}.
Even though the form of the duality as expressed in these references is
restricted to $\Z{2}\times\Z{2}'$ orbifolds, the important idea that the
spectral flow map can be used to relate different \twozero{} models is
much more general. It is the purpose of this paper to explain the
details of this mapping and the exact relationship between the mapped
models.

The outline of this paper is as follows: In section \ref{sec:svd}
we discuss the spinor-vector duality in the fermionic $\Z{2}\times \Z{2}$
heterotic-string orbifolds.
Understanding how the duality operates in the free fermionic constructions
hints at how similar dualities may work in the case of interacting CFTs.
In sections \ref{sec:spectral flow} and \ref{sec:simple currents} we review 
the definition of the spectral flow and the simple current formalism which will 
allow us to construct \twozero{} models from a generic $(2,2)$ model. In section \ref{sec:idea}
we explain how the spectral flow induces a map between different \twozero{} models and in section \ref{sec:mapping} we analyze
the consequences of this idea. Section \ref{sec:example} provides an example of how
to use the derived results. We conclude with a brief discussion and possible generalizations
in sections \ref{sec:generalizations} and \ref{sec:conclusions}.

\section{The spinor-vector duality case}\label{sec:svd}

In this section we outline the spinor-vector duality in the case
of the fermionic $\Z{2}\times \Z{2}$ heterotic-string orbifolds. 
The discussion will provide the guide for exploring 
similar symmetries in models with an interacting internal CFT. 
The presentation here will be qualitative and further
technical details are given in the references. 

In the free fermionic formulations of the compactified string \cite{fff}
all the internal degrees of freedom are represented in terms of
free world-sheet fermions. Therefore, in this formulation
the internal compactified dimensions are represented in terms 
of an internal CFT with vanishing interactions. Additionally, the
well known relations between two dimensional fermions and bosons
entail that the free fermionic formulation is equivalent to 
a free bosonic formulation, {\it i.e.\ }to toroidal orbifolds.

A string vacuum in the free fermionic formulation is defined
in terms of boundary condition basis vectors and the Generalized
Gliozzi-Scherk-Olive (GGSO) projection coefficients of the one-loop partition function
\cite{fff}. The gauge symmetry is generated by spacetime 
vector bosons that arise from the untwisted as well as the twisted 
sectors. The spacetime vector bosons arising in the twisted sectors enhance the
untwisted gauge group factors under which they are charged. 
Specific enhancements depend on the states that remain in the 
physical spectrum after application of the GGSO projections. 
Similarly, the matter states in the free fermionic models
are obtained from the untwisted and twisted sectors.
The spinor-vector duality in the free fermionic vacua operates on the 
matter states in the twisted sectors.

The free fermionic vacua correspond to $\Z{2}$ and $\Z{2}\times \Z{2}$
orbifolds at enhanced symmetry points in the moduli space \cite{z2xz2}.
In this section we review the spinor-vector duality in $\Z{2}$ orbifolds.
By doing this we recap the ingredients that are needed for the generalization to 
interacting internal CFTs. The simplest realization of the spinor-vector
duality is in the case of a single $\Z{2}$ orbifold acting on the
$E_8\times E_8$ heterotic-string compactified on a generic six torus.
Taking for simplicity the internal torus as a product of six circles with 
radii $R_i$, the partition function (omitting the contribution from the spacetime bosons) reads
\beq
{ Z}_+ = ( V_8 - S_8) \, \left( \sum_{m,n} \Lambda_{m,n}
\right)^{\otimes 6}\, \left(  \overline O _{16} + \overline S_{16} \right)
\left( \overline O _{16} +  \overline S_{16} \right)\,,
\eeq
where as usual, for each circle,
\beq
p_{\rm L,R}^i = \frac{m_i}{R_i} \pm \frac{n_i R_i}{\alpha '} \,,
~~~~~{\rm and}~~~~~
\Lambda_{m,n} = \frac{{q^{\frac{\alpha '}{4} 
p_{\rm L}^2}} \, \bar q ^{\frac{\alpha '}{4} p_{\rm R}^2}}{|\eta|^2}\,,
\eeq
and we have written ${Z}_{+}$ in terms of level-one
${\rm SO} (2n)$ characters (see for instance \cite{Angelantonj2002})
\beqn
O_{2n} &=& \frac{1}{2} \left( \frac{\theta_3^n}{\eta^n} +
\frac{\theta_4^n}{\eta^n}\right) ~~~~\,,
~~~~~~~~~~~~
V_{2n} ~=~ \frac{1}{2} \left( \frac{\theta_3^n}{\eta^n} -
\frac{\theta_4^n}{\eta^n}\right),
\nonumber \\
S_{2n} &=& \frac{1}{2} \left( \frac{\theta_2^n}{\eta^n} +
i^{-n} \frac{\theta_1^n}{\eta^n} \right) \,,
~~~~~~~~~~~
C_{2n} ~=~ \frac{1}{2} \left( \frac{\theta_2^n}{\eta^n} -
i^{-n} \frac{\theta_1^n}{\eta^n} \right).
\nn
\eeqn
We next apply the orbifold projections 
\beqn
\Z{2}~:~g & = &(-1)^{F_{1}+F_{2}}\delta \,, \label{deltaorbifold}\\
\Z{2}^\prime ~:~{g^\prime} & = &~(x_{4},x_{5},x_{6},x_7,x_8,x_9)
~\longrightarrow~
(-x_{4},-x_{5},-x_{6},-x_7,+x_8,+x_9)\,.\nonumber
\eeqn
$F_1$ and $F_2$ in (\ref{deltaorbifold}) flip the sign in the
spinorial representations of $SO(16)_1$ and $SO(16)_2$, generated
by ${\xi_1}=\{{\bar\psi}^{1,\cdots,5}, {\bar\eta}^{1,2,3}\}$
and ${\xi_2}=\{{\bar\phi}^{1,\cdots,8}\}$ respectively, and $\delta$
shifts the compact $X^9$ coordinate by half of its period, {\it i.e.}
\begin{equation} 
\delta ~:~ X^9 \rightarrow X^9 +\pi R^9 \qquad \Rightarrow \qquad
\Lambda_{m,n} ~\rightarrow~ (-1)^m\Lambda_{m,n}.\label{shift1}
\end{equation}
The $\Z{2}$ projection in (\ref{deltaorbifold})
breaks the $E_8\times E_8$ gauge group to $SO(16)\times SO(16)$ 
and preserves $N=4$ spacetime supersymmetry.
The additional $\Z{2}^\prime$ projection twists the compactified coordinates
and preserves only $N=2$ spacetime supersymmetry. Its generator
$g^\prime$ reverts the sign of four internal coordinates $X^i$, $i=4,5,6,7$ 
and simultaneously breaks one $SO(16)$ to $SO(12)\times SO(4)$. 
 
The action of the $\Z{2}\times \Z{2}^\prime$ projections on $Z_+$ is implemented
by taking
\beq 
Z_-=
\frac{\left(1+g\right)}{2} 
\frac{\left(1+g^\prime\right)}{2}
Z_+\,.
\label{tendprojection}
\eeq
The ten-dimensional SO(8) little group is broken to $SO(4) \times SO(4)$.
At the same time, the first $SO(16)$ gauge
group factor is broken into $SO(12) \times SO(4)$. As a
result, the one-loop partition function can be written in terms of the
spacetime characters,
\beqn
Q_0 & = & V_4O_4-S_4S_4,~~~~~~~~~~~~~Q_V = V_4O_4- C_4C_4,\nn\\
Q_S & = & O_4C_4-S_4O_4,~~~~~~~~~~~~Q_C = V_4S_4- C_4V_4.\nn
\eeqn
There are two independent orbits in the partition function and hence
one discrete torsion. The full partition function is given by
\beq
Z_-=Z_{untwisted}+Z_{g}+Z_{g^\prime}+Z_{gg^\prime}.
\eeq
It consists of the untwisted sector and 
the three twisted sectors $g$, $g^\prime$ and $gg^\prime$.
The untwisted sector gives rise to the vector bosons 
that generate the four dimensional gauge group, whereas 
the sectors $g$ and $gg^\prime$ give rise to massive states. 
To note the spinor-vector duality it is sufficient to 
focus on the states arising from the twisted sector $g^\prime$. 
Summation over the GGSO projections in this 
sector produces the following terms in the partition function: 
\beqn
Z_{g'}~& & = \nonumber\\
& &
{1\over 2}
\left(
\left\vert{{2\eta}\over\theta_4}\right\vert^4
+
\left\vert{{2\eta}\over\theta_3}\right\vert^4
\right)
\Lambda_{p,q}
\left[
P^{01}_+\Lambda_{m,n}
\left(
Q_S \left(
{\overline V}_{12}{\overline C}_{4}{\overline O}_{16}
+
{\overline S}_{12}{\overline O}_{4}{\overline S}_{16}
\right) 
 \right. \right. \nonumber\\
& & 
   \left.   \left.
~~~~~~~~~~~~~~~~~~~~~~~~~~~~+~~~~~~~~~~~~~~~~
Q_C
\left(
{\overline O}_{12}{\overline S}_{4}{\overline O}_{16}
+
{\overline C}_{12}{\overline V}_{4}{\overline S}_{16}
\right)
\right)
\right.
\nonumber\\
& &  
~~~~~~~~~~~~~~~~~~~~~~~~~~~~~+~~
\left.
P^{01}_-\Lambda_{m,n}
\left(
Q_S \left(
{\overline S}_{12}{\overline O}_{4}{\overline O}_{16}
+
{\overline V}_{12}{\overline C}_{4}{\overline S}_{16}
\right)
\right. \right. \nonumber\\
& & 
\left. \left.
~~~~~~~~~~~~~~~~~~~~~~~~~~~~+~~~~~~~~~~~~~~~~
Q_C
\left(
{\overline O}_{12}{\overline S}_{4}{\overline S}_{16}
+
{\overline C}_{12}{\overline V}_{4}{\overline O}_{16}
\right)\right)
\right]
\nonumber
\\
& &
+{1\over 2}
\left(
\left\vert{{2\eta}\over\theta_4}\right\vert^4
-
\left\vert{{2\eta}\over\theta_3}\right\vert^4
\right)
\Lambda_{p,q}
\left[
P^{01}_+\Lambda_{m,n}
\left(
Q_S \left(
{\overline O}_{12}{\overline S}_{4}{\overline O}_{16}
+
{\overline C}_{12}{\overline V}_{4}{\overline S}_{16}
\right) 
 \right. \right. \nonumber\\
& & 
 \left.   \left.
~~~~~~~~~~~~~~~~~~~~~~~~~~~~+~~~~~~~~~~~~~~~~
Q_C
\left(
{\overline V}_{12}{\overline C}_{4}{\overline O}_{16}
+
{\overline S}_{12}{\overline O}_{4}{\overline S}_{16}
\right)
\right)
\right. 
\nonumber\\
& &  
~~~~~~~~~~~~~~~~~~~~~~~~~~~~~+~~
\left.
P^{01}_-\Lambda_{m,n}
\left(
Q_S \left(
{\overline O}_{12}{\overline S}_{4}{\overline S}_{16}
+
{\overline C}_{12}{\overline V}_{4}{\overline O}_{16}
\right)
 \right. \right. \nonumber\\
& & 
 \left. \left.
~~~~~~~~~~~~~~~~~~~~~~~~~~~~+~~~~~~~~~~~~~~~~
Q_C
\left(
{\overline S}_{12}{\overline O}_{4}{\overline O}_{16}
+
{\overline V}_{12}{\overline C}_{4}{\overline S}_{16}
\right)\right)
\right],
\label{partifun}
\eeqn
where we defined 
$P_\pm^{01}$ as 
\beq
P_{\pm}^{01}={{1\pm\epsilon_1(-1)^m}\over2}.
\label{ppm}
\eeq 

The spinor-vector duality transformation is transparent
in the partition function (\ref{partifun}). Massless states arise from
the untwisted sector and the $g'$-twisted sector. 
The internal winding modes in the $g$ and $gg'$-twisted sectors 
are shifted by $1/2$. The states in these two sectors are therefore
massive. The untwisted sector gives rise to spacetime
vector bosons that generate the $SO(12)\times SO(4)\times SO(16)$ 
gauge symmetry and to scalar multiplets that transform in the
bi-vector representation of $SO(12)\times SO(4)$. Examining the
$g'$-twisted sector reveals how the spinor-vector
duality operates. Massless states arise for vanishing internal momentum
and winding modes, {\it i.e.\ }$m=n=0$. Depending on the choice of the 
discrete torsion $\epsilon_1=\pm1$, vanishing lattice modes will 
therefore arise from $P_+^{01}\Lambda_{m,n}$ or $P_-^{01}\Lambda_{m,n}$, 
{\it i.e.} 
\beqn
& &\epsilon_1=+1~~\Rightarrow ~~~P^{01}_+\Lambda_{m,n}=\Lambda_{2m,n} 
~~~{\rm and~~~}          P^{01}_-\Lambda_{m,n}=\Lambda_{2m+1,n}\ , \nn\\
& &\epsilon_1=-1~~\Rightarrow ~~~P^{01}_-\Lambda_{m,n}=\Lambda_{2m,n} 
~~~{\rm and~~~}          P^{01}_+\Lambda_{m,n}=\Lambda_{2m+1,n}\ . \nn
\eeqn
It follows from the $q$-expansion of the $\theta$ functions
that in the case with $\epsilon_1=+1$ the zero lattice modes
attach to $Q_S{\overline V}_{12}{\overline C}_{4}{\overline O}_{16}$,
which produces two massless $N=2$ hypermultiplets in the 
$\bf{12}$ vector representation of $SO(12)$, whereas
in the case with  $\epsilon_1=-1$ the zero lattice modes
attach to $Q_S{\overline S}_{12}{\overline O}_{4}{\overline O}_{16}$,
which produces a massless $N=2$ hypermultiplet in the
$\bf{32}$ spinorial representation. It is further noted from 
(\ref{partifun}) that
in the case with $\epsilon_1=+1$ the term 
$Q_S{\overline O}_{12}{\overline S}_{4}{\overline O}_{16}$
gives rise to eight additional states from the first 
excited modes of the twisted lattice. Hence, the total number 
of degrees of freedom $32=12\cdot2 +4\cdot 2$ is preserved under 
the duality map. 

The realization of the spinor-vector duality in this model provides a simple
example where its origins can be explored and generalized to cases with 
interacting world-sheet CFTs. In the toroidal case, since all the data 
of the compactification is encoded in the toroidal background parameters
and the orbifold action on them, it is anticipated that the spinor-vector 
duality is realizable in terms of a continuous or discrete map between two 
sets of background parameters. Indeed, in ref. \cite{Faraggi2011}
it was shown that the spinor-vector duality map is realized in terms of
a continuous interpolation between two Wilson lines. The continuous 
interpolation, rather than a discrete transformation, is particular to the
cases that preserve $N=2$ spacetime supersymmetry, {\it i.e.\ }when
 a single $\Z{2}$ twist is acting  on the internal torus. In this 
case the moduli fields that enable the continuous interpolation
exist in the spectrum and are not projected. In the compactifications
with $N=1$ spacetime supersymmetry, these moduli fields are
projected out. Therefore, in the $N=1$ cases the spinor-vector
duality map is discrete. 

The spinor-vector duality can be regarded as a direct consequence of
the breaking of the world-sheet supersymmetry on the bosonic side of
the heterotic-string from $N=2$ to $N=0$,  {\it i.e.\ }from $(2,2)$
world-sheet supersymmetry to \twozero. In the $(2,2)$ case the gauge
symmetry is enhanced to $E_6$ (or $E_7$). In this case the spinor and
vector representations of $SO(10)\times U(1)$ (or $SO(12)\times SU(2)$)
are embedded in the single $\bf{27}$ (or $\bf{56}$) representation of $E_6$ (or $E_7$).
The breaking of the $(2,2)$ world-sheet supersymmetry to \twozero{} results 
in the reduction of the enhanced gauge symmetry, by projecting out the spinorial 
components of the adjoint representation in its decomposition under the
corresponding $SO(2n)$ subgroup.
At the same time the matter multiplets are split into
the spinorial and vectorial components. The GSO projections may
retain either the spinorial or the vectorial representation in the massless
spectrum. The spinor-vector duality is then induced by the spectral
flow operator. The generalization to interacting internal CFTs can therefore
proceed along the following lines. We can start with a generic 
compactification with $(2,2)$ world-sheet supersymmetry,
and subsequently break the $N=2$ world-sheet supersymmetry on
the bosonic side to $N=0$. There ought to be choices of the breaking that
result in different models that are related by the spectral flow operator.

We can illustrate the spinor-vector duality in terms of a spectral
flow operator by considering the boundary condition basis
vectors \cite{Faraggi2008} in eq. (\ref{neq2basis}): 
\begin{eqnarray}
v_1=S&=&\{\psi^\mu,\chi^{1,\dots,6}\},\nn\\
v_{1+i}=e_i&=&\{y^{i},\omega^{i}|\bar{y}^i,\bar{\omega}^i\}, \
i=1,\dots,6,\nn\\
v_{8}=z_1&=&\{\bar{\phi}^{1,\dots,4}\},\nn\\
v_{9}=z_2&=&\{\bar{\phi}^{5,\dots,8}\},\nn\\
v_{10}=z_3&=&\{\bar{\psi}^{1,\dots,4}\},\nn\\
v_{11}=z_0&=&\{\bar{\eta}^{0,1,2,3}\},\nn\\
v_{12}=b_1&=&\{\chi^{34},\chi^{56},y^{34},y^{56}|\bar{y}^{34},
\bar{y}^{56},\bar{\eta}^0,\bar{\eta}^{1}\},\label{neq2basis}
\end{eqnarray}
where the vector ${\bf1}= \{\psi^\mu,\
\chi^{1,\dots,6},y^{1,\dots,6}, \omega^{1,\dots,6}| 
~\bar{y}^{1,\dots,6},\bar{\omega}^{1,\dots,6},
\bar{\eta}^{1,2,3},
\bar{\psi}^{1,\dots,5},\bar{\phi}^{1,\dots,8}\}$ 
is obtained as the linear combination 
${\bf1} =S+{\sum_ie_i}+z_0+z_1+z_2+z_3$. 
In (\ref{neq2basis}) we used
the usual notation of the free fermionic formalism \cite{fff}.
The gauge group generated by vector bosons arising in the $0$-sector
is $SO(8)\times SO(8)\times SO(8)\times SO(8)$. 
The gauge symmetry may be enhanced by vector bosons arising from 
nine additional purely anti-holomorphic sets given by: 
\beqn
G=\{& z_0,z_1,z_2,z_3, \nonumber\\
    & z_0+z_1,z_0+z_2,z_0+z_3,z_1+z_2,z_1+z_3,z_2+z_3~\}. 
\label{gaugebosons}
\eeqn
The basis vector $b_1$ reduces
the $N=4\rightarrow N=2$ spacetime supersymmetry
and the untwisted gauge symmetry to 
$SO(8)\times SO(4)\times SO(4)\times SO(8)\times SO(8)$. 
Additionally, it gives rise to the twisted sector, which produces 
matter states charged under the four dimensional gauge group.
The sixteen sectors $B_1^{pqrs}=b_1+pe_3+qe_4+re_5+se_6$, with 
$p,q,r,s\in\{0,1\}$, correspond to the sixteen fixed points of
the non-freely acting $\Z{2}$ orbifold. 

For specific choices of the GGSO projection coefficients the 
gauge group is enhanced. The vector bosons arising 
from the sector $z_3$ may enhance the $SO(8)\times SO(4)\times SO(4)$
symmetry to $SO(12)\times SO(4)$, which may be enhanced further to 
$E_7\times SU(2)$. In the case of $E_7$ both $z_3$ and $z_0$ are generators
of the $E_7$ gauge group. In the case of $SO(12)$ the matter
representations are obtained from the following sectors: the two sectors $B^{pqrs}_1$ and 
$B^{pqrs}_1+z_3$ give the vectorial $\bf{12}$ representation and the two 
sectors $B^{pqrs}_1+z_0$ and $B^{pqrs}_1+z_3+z_0$ the spinorial $\bf{32}$.

For appropriate choices of the GGSO phases either the spinorial or the vectorial representations
from a given sector are retained in the spectrum. If both the spinorial
and the vectorial states are retained in a given sector, the $SO(12)\times
SU(2)$ symmetry is necessarily enhanced to $E_7$. We note therefore
that it is precisely the basis vector $z_0$ that acts as the spectral 
flow operator. For an appropriate choice of the phases it acts as a
generator of $E_7$, whereas when the $E_7$ symmetry is broken to 
$SO(12)\times SU(2)$, coupled with appropriate 
mapping of the GGSO projections,
the spinor-vector duality map is induced. Examining the basis 
vectors in (\ref{neq2basis}) we see that $z_0$ is precisely the
mirror of the basis vector $S$, which is the spacetime
supersymmetry generator on the fermionic side of the heterotic-string. 
Hence, $S$ is an operator of the \lefto-moving $N=2$ world-sheet 
supersymmetry, whereas $z_0$ is an operator of the world-sheet 
supersymmetry on the bosonic side. 

An important feature of the  $\Z{2}\times\Z{2}'$ models is that the spectral flow
operator is of order two, {\it i.e.\ }the sector $2z_0$ is identified with the untwisted sector.
This leads to two different models, as explained above, related via the spinor-vector
duality. In this paper we generalize these ideas to arbitrary internal RCFTs.
For these, the spectral flow operator will generically be of order greater than two leading
naturally to a bigger family of models. In the following sections we explain how these models
are related in the most general case.  

\section{The spectral flow}\label{sec:spectral flow}
To handle the most general case in what follows, we will be slightly changing our notation from the one used in the previous section and in the free fermionic language. Our starting point here is generic (2,2) heterotic models with an internal CFT with c=9. The standard examples of interacting constructions are the Gepner models\cite{Gepner} in which the internal CFT is a product of minimal models, but all our arguments are completely general. A general state in such a model is of the form:
\beq\label{eq:1}
\Phi_\L\otimes\Phi_\R
\eeq
and the \righto-moving part which we wish to focus on is of the form
\beq\label{eq:2}
\Phi_\R=(w)(h,Q)(p),
\eeq
where $w$ is an $SO(10)$ weight $(o, v, s, c)$ and $p$ an $E_8$ weight. The appearance of the $SO(10)$ and $E_8$ weights is because of the bosonic string map which is used to construct a modular invariant heterotic-string theory from a type II theory. It replaces the $\widehat{so}(2)_1$ Kac-Moody algebra with an $\widehat{so}(10)_1\times (\widehat{e_8})_1$ one\cite{books}.

The mass formula is
\beqn
\frac{\alpha'M_R^2}{2}&=&h_{\text{TOT}}-\frac{c}{24}\nn\\
&=&\frac{w^2}{2}+h+\frac{p^2}{2}+N_R-1\ ,\label{MR}
\eeqn
where we have used the fact that $c=24$ for the bosonic string and we have also included the contribution $N_R$ from the oscillators corresponding to the spacetime bosons.

By definition a CFT is said to have $N=2$ world-sheet supersymmetry if it includes four fields:
\beqn
T(z)&=&\sum_{n\in\Z{}}L_nz^{-n-2}\ ,\\
G^{\pm}(z)&=&\sum_{n\in\Z{}}G^{\pm}_{n\pm a}z^{-n-\frac{3}{2}\mp a}\ ,\\
J(z)&=&\sum_{n\in\Z{}}J_nz^{-n-1}\ ,
\eeqn
that satisfy the algebra\cite{books}:
\beqn
\left[L_m,L_n\right]&=&(m-n)L_{m+n}+\frac{c}{12}(m^3-m)\delta_{m+n,0}\ ,\nn\\
\left[L_m, G^\pm_{n\pm a}\right]&=&(\frac{m}{2}-n\mp a)G^\pm_{m+n\pm a}\ ,\nn\\
\left[L_m, J_n\right]&=& -n J_{m+n}\ ,\nn\\
\left[J_m, J_n\right]&=& \frac{c}{3}m\delta_{m+n,0}\ ,\nn\\
\left[J_m, G^\pm_{n\pm a}\right]&=&\pm G^\pm_{m+n\pm a}\ ,\nn\\
\{G^+_{m+a},G^-_{n-a}\}&=&2 L_{m+n}+(m-n+2a)J_{m+n}+\frac{c}{3}\Big((m+a)^2-\frac{1}{4}\Big)\delta_{m+n,0}\ ,\nn\\
\{G^+_{m+a},G^+_{n+a}\}&=&\{G^-_{m-a},G^-_{n-a}\}=0\ ,
\eeqn
where $a$ is a real parameter that describes how the fermionic superpartners $G^{\pm}$ of $T$ transform:
\beq
G^{\pm}(e^{2\pi i}z)=-e^{\mp2\pi i a}G^{\pm}(z).
\eeq
The algebras for $a$ and $a+1$ are isomorphic. $a\in\Z{}$ corresponds to the R sector and $a\in\Z{}+\frac1 2$ corresponds to the NS sector. A state is completely described by the eigenvalues $h$ (called the conformal dimension) and $Q$ (called the $U(1)$ charge) of the operators $L_0$ and $J_0$ that form the Cartan  subalgebra:
\beq
\bt{\phi}=\bt{h,Q}.
\eeq
 We also note that the algebra is invariant under the following transformation which is known as the \emph{spectral flow}:
\beqn
L_n^\eta&=&L_n+\eta J_n+\frac{c}{6}\eta^2\delta_{n,0}\ ,\nn\\
G^{\eta \pm}_{n\pm a}&=& G^{\eta \pm}_{n\pm(a+\eta)}\ ,\nn\\
J^\eta_n&=&J_n+\frac{c}{3}\eta\delta_{n,0}.
\eeqn
This also implies the existence of a \emph{spectral flow operator} $U_\eta$ that acts on states in the following way:
\beq
U_\eta\bt{h,Q}=\bt{h_\eta,Q_\eta}={\big| h-\eta Q+\frac{\eta^2 c}{6},Q-\frac{c \eta}{3} \big\rangle}.
\eeq

Of particular interest are the states
\beq
\big| \frac{3}{8},\pm\frac{3}{2} \big\rangle_{\text{R}}=U_{\mp\frac{1}{2}}{\bt{0,0}}_\text{NS}\ ,
\eeq
because they can be combined with the $s$ and $c$ weight vectors of $SO(10)$ with the smallest possible length to give massless states. Indeed, such vectors are of the form 
\beq
w=(\pm\frac{1}{2},\pm\frac{1}{2},\pm\frac{1}{2},\pm\frac{1}{2},\pm\frac{1}{2})
\eeq
and have $w^2=\frac{5}{4}$. An even number of minus signs corresponds to $s$ and an odd number of minus signs to $c$. We then note from (\ref{MR}) that whenever the internal CFT has $N=2$ world-sheet supersymmetry the states
\beq\label{beta0}
\pm\beta_0=(\pm c)(\frac3 8,\pm\frac3 2)(0).
\eeq
will be part of the massless spectrum. These states describe gauge bosons in the \sixt{} and \sixtbar{} of $SO(10)$ and, in conjunction with the $U(1)$ symmetry of the $N=2$ algebra, they extend $SO(10)$ to $E_6$. This proves our previous claim that the $N=2$ superconformal algebra on the bosonic sector is associated with  $E_6$ gauge symmetry.
The states in (\ref{beta0}) are an extension of the spectral flow operator of the internal CFT. We call these states the spectral flow operator as well.

\section{The simple current formalism}\label{sec:simple currents}
Since we already started from a (2,2) model, there will be a modular invariant partition function (MIPF) describing it. It will be of the form
\beq
Z[\tau,\bar\tau]=\sum_{i,j} \chi_i(\tau)M_{ij}\chi_{j}(\bar\tau),
\eeq
where $\chi_i$ are the characters of the chiral algebra and $M_{ij}$ a modular invariant. For our examples, we take this to be the partition function of the usual Gepner models, \textit{i.e.\ }after the projections of the universal simple currents $\beta_0$ and $\beta_i$ have been applied to ensure spacetime supersymmetry\cite{Gepner}. Nevertheless, the approach is very general and valid whenever the simple current method can be used to construct modular invariants. This includes any RCFT and potentially some non-rational CFTs in which the chosen simple current defines a finite orbit as well. To avoid this complication we restrict ourselves to RCFTs through this paper.

As explained in the introduction we are not interested in the $(2,2)$ models \textit{per se} but rather in the \twozero{} that we get after breaking the $E_6$ symmetry on the \righto. A consistent and modular invariant \twozero{} model can be derived from a $(2,2)$ model through the simple current construction\cite{Schellekens1990,Schellekens1989b}. This is the same as the beta method for Gepner models and it practically amounts to orbifolding the original $(2,2)$ model. The result is that states not invariant under the action of the simple current are projected out and new states appear in twisted sectors. We will use both notations $J$ and $\beta$ for a simple current\footnote{Using multiplicative notation for the action of $J$ and additive notation for the action of $\beta$.} and we will focus on simple currents that break $E_6$ on the \righto{} to $SO(10)$. The MIPF for the resulting model is then
\beq
Z[\tau,\bar\tau]=\sum \chi_i(\tau)M_{ik}M_{kj}(J)\chi_{j}(\bar\tau),
\eeq
where
\beq\label{eqn:SCMI}
M_{kj}(J)=\frac 1 N\sum_{n=1}^{N_J}\delta({\Phi_k,J^n\Phi_j})\delta_{\mathbb{Z}}(Q_J(\Phi_k)+\frac{n}{2}Q_J(J))
\eeq
is called a simple current modular invariant (SCMI) and $N$ is a normalization constant ensuring that the vacuum only appears once. In practical terms, the above formula means that:
\begin{enumerate}[i)]
\item Only states whose \lefto{} part is connected to the \righto{} through $J$ will appear in the partition function, \textit{i.e.\ }states with $\Phi_\L=J^n\Phi_\R=\Phi_\R+n\beta$. This defines the $n$-th $J$-twisted sector.  
\item Only states invariant under the projection will appear in the partition function. This is expressed in the constraint $Q_J(\Phi)+\frac{n}{2}Q_J(J)\in \mathbb{Z}$. $Q_J$ is called the monodromy charge and is defined as
\beq\label{eq:monodromycharge}
Q_J(\Phi)=h(\Phi)+h(J)-h(J\Phi)\quad \mod1.
\eeq

The easiest way to see that this is the appropriate condition for invariance under the $J$ projection is to note that the monodromy charge is conserved modulo 1 in operator products and thus implies the existence of a phase symmetry $\Phi\rightarrow e^{-2\pi i Q_J(\Phi)}\Phi$. This induces a cyclic group of order $N_J$. $N_J$ is called the order of $J$ and it can also be proven that $Q_J(\Phi)$ is quantized in units of $1/N_J$ \cite{Schellekens1989b}.
\end{enumerate}
The definition (\ref{eq:monodromycharge}) is for any general RCFT. For Gepner models, where $\Phi=(w_{\Phi})(\vec{l}_{\Phi},\vec{q}_{\Phi},\vec{s}_{\Phi})(p_{\Phi})$ and $J=(w_J)(\vec{l}_{J},\vec{q}_{J},\vec{s}_{J})(p_{J})$, it takes the explicit form:
\beq\label{eq:monodromychargeinGepner}
Q_J(\Phi)=-w_J\cdot w_{\Phi}-p_J\cdot p_{\Phi}+\sum_{i=1}^{r}\left(\frac{-l^i_{\Phi}l^i_J+q^i_{\Phi}q^i_J}{2(k_i+2)}-\frac{s^i_{\Phi}s^i_J}{4}\right).
\eeq
In this form it is easy to see that
\beq
Q_{\beta}(\Phi)=Q_{\Phi}(\beta)\quad \mbox{and}\quad Q_{\beta_1+\beta_2}(\Phi)=Q_{\beta_1}(\Phi)+Q_{\beta_2}(\Phi),
\eeq
\textit{i.e.\ }the monodromy charge is symmetric and linear with respect to its arguments. These properties are true in general\cite{Schellekens1989b}.

Another thing to note is that if $J$ and $J'$ are simple currents then $JJ'$ is a simple current as well. In fact, we can generalize (\ref{eqn:SCMI}) to the case where we orbifold by $J_1,\cdots,J_i,\cdots$ simultaneously. To simplify the notation let $\vec{n}$ label the twisted sectors and define 
$$[\vec{n}]k\equiv J_1^{n_1}\cdots J_i^{n_i}\cdots\Phi_k\equiv\Phi_k+\sum_i n_i \beta_i.$$
Then the most general SCMI is \cite{Kreuzer1994}:
\beq\label{eqn:SCMItorsion}
M_{k,[\vec{n}]k}=\frac 1 N\prod_i\delta_{\mathbb{Z}}(Q_{J_i}(\Phi_k)+X_{ij}n^j).
\eeq
The matrix $X$ is defined modulo 1 and its elements are quantized as $X_{ij}=\frac{n_{ij}\in\mathbb{Z}}{\gcd(N_i,N_j)}$. It also satisfies $X_{ij}+X_{ji}=Q_{J_i}(J_j)$. This fixes its symmetric part completely. The remaining freedom in choosing the antisymmetric part corresponds to discrete torsion\cite{Kreuzer1994}.

\section{Outline of the idea}\label{sec:idea}
We start with a particular simple current $J$. Any $J$ would do, but for the reasons explained in the introduction the simple currents that we have in mind will break $E_6$, thus giving a \twozero{} model. We call the \twozero{} model that is derived this way ${\cal{M}}_0$. We also know that $J_0$ ($\beta_0$) is generically a simple current of every $(2,2)$ model since it is the spectral flow operator that enhances the symmetry to $E_6$ on the \righto. This naturally defines a whole family of models $\{\cal{M}\}_\alpha$ that are derived through the simple currents $J$, $J_0$ and linear combinations of them with and without discrete torsion. 

The task of examining how the spectra of these models are related to each other is very fascinating and daunting at the same time. We will not attempt to carry out the analysis in its full generality here. Instead, we will restrict ourselves to the more modest goal of explaining how the mapping induced by the spectral flow $J_0$ ($\beta_0$) works.

\section{Mapping induced by the spectral flow}\label{sec:mapping}
Here we focus on the family of models  ${\cal{M}}_0,\cdots,{\cal{M}}_m$ that are derived through the simple currents $J, JJ_0, \cdots, JJ_0^m$ or equivalently $\beta, \beta+\beta_0,\cdots,\beta+m\beta_0$. This family will have $N_{\beta_0}$ members where  $N_{\beta_0}$ is the order of  $\beta_0$. Our goal is to study how the massless spectra in these models are related. To that end, we take a closer look at the model ${\cal{M}}_m$.

We start by examining the untwisted sector\footnote{Here and in what follows untwisted sector means untwisted with respect to the simple current that defines the model,  {\it i.e.\ }states with $n=0$ in (\ref{eq:n-twisted sector}). The states might be twisted with respect to other simple currents that were present in the original (2,2) model but this does not affect our argument.}. Massless states in the original $(2,2)$ model will also belong to the ${\cal{M}}_m$ model if they survive the invariance projections. Note that
\beq
Q_{\beta+m\beta_0}(\Phi)=Q_{\beta}(\Phi)+mQ_{\beta_0}(\Phi)=Q_{\beta}(\Phi)\ \mod 1,
\eeq
where in the last step we used the fact that $Q_{\beta_0}(\Phi)\in\Z{}$ because $\Phi$ belongs to the original $(2,2)$ model. This proves that $Q_{\beta+m\beta_0}(\Phi)\in\Z{}\Leftrightarrow Q_{\beta}(\Phi)\in\Z{}$ and therefore the untwisted sectors of every model in the $\cal{M}$ family are identical.

Let us now consider the twisted sectors. Note that models ${\cal{M}}_{m_1}$ and ${\cal{M}}_{m_2}$ will  in general have a different number of twisted sectors since $\beta+m_1\beta_0$ and   $\beta+m_2\beta_0$ will be of different order. Let us analyze the $n$-twisted sector of the ${\cal{M}}_m$ model. A very useful formula can be found by rearranging (\ref{eq:monodromycharge}) as 
\beqn\label{eq:hmbeta}
h(\Phi+\beta)&=&h(\Phi)+h(\beta)-Q_{\beta}(\Phi)\ ,\nonumber\\
\mbox{and by induction:}\quad h(\Phi+m\beta)&=&h(\Phi)+m h(\beta)-m Q_{\beta}(\Phi)-\frac{m(m-1)}{2}Q_{\beta}(\beta)\ ,
\eeqn
where the equations are understood mod $1$. Massless states in the $n$-twisted sector of ${\cal{M}}_m$ are of the form
\beq\label{eq:n-twisted sector}
\Phi_\L\otimes(\tilde\Phi_\L+n(\beta+m\beta_0))\ ,
\eeq
where this time we have written the tilde explicitly to remind us that we have applied the bosonic string map. In the notation of equation (\ref{eq:2}) this is simply\cite{Gepner}:
\beq
\tilde\Phi_\L=\Phi_\L+(v)(0,0)(0).
\eeq

The massless condition gives
\beq\label{eq:massless}
h(\Phi_\L)=\frac1 2,\quad h(\tilde\Phi_\L)=1 \quad\mbox{and}\quad h(\tilde\Phi_\L+n\beta+nm\beta_0)=1.
\eeq
Furthermore, as explained before and as can be seen from (\ref{eqn:SCMI}), the states must also satisfy the invariance condition
\beq
Q_{\beta+m\beta_0}(\tilde\Phi_\L)+\frac{n}{2}Q_{\beta+m\beta_0}(\beta+m\beta_0)\in\Z{}.
\eeq
Using linearity of the monodromy charge and the fact that $Q_{\beta_0}(\tilde\Phi_\L)\in\Z{}$ and $\ Q_{\beta_0}(\beta_0)\in2\Z{}$ because $\tilde\Phi_\L$ and $\beta_0$ belonged to the massless spectrum of the original $(2,2)$ model, the invariance condition becomes
\beq\label{eq:invariant}
Q_{\beta}(\tilde\Phi_\L)+\frac{n}{2}Q_{\beta}(\beta)+mn Q_{\beta_0}(\beta)\in\Z{}.
\eeq
We can also further manipulate (\ref{eq:massless}) to derive another condition. Bearing in mind that in what follows all the calculations are mod $1$, we get:
\beqn
0=1&=&h(\tilde\Phi_\L+n\beta+nm\beta_0)\nonumber\\
&\stackrel{(\ref{eq:hmbeta})}{=}&h(\tilde\Phi_\L+n\beta)+\underbrace{nmh(\beta_0)}_{\in\Z{}}-\underbrace{nmQ_{\beta_0}(\tilde\Phi_\L)}_{\in\Z{}}-n^2mQ_{\beta_0}(\beta)-\underbrace{\frac{nm(nm-1)}{2}}_{\in\Z{}}\underbrace{Q_{\beta_0}(\beta_0)}_{\in\Z{}}\nonumber\\
&=& h(\tilde\Phi_\L+n\beta)-n^2mQ_{\beta_0}(\beta)\nonumber\\
&\stackrel{(\ref{eq:hmbeta})}{=}&  \underbrace{h(\tilde\Phi_\L)}_{=1=0}+nh(\beta)-nQ_{\beta}(\tilde\Phi_\L)-\frac{n(n-1)}{2}Q_{\beta}(\beta)-n^2mQ_{\beta_0}(\beta)\nonumber\\
&\stackrel{(\ref{eq:invariant})}{=}& nh(\beta)+\frac{n}{2}Q_{\beta}(\beta)
\eeqn
Or in other words,
\beq\label{eq:condition2}
n\Big(h(\beta)+\frac{1}{2}Q_{\beta}(\beta)\Big)\in\Z{}.
\eeq

Equations (\ref{eq:invariant}) and (\ref{eq:condition2}) are the main results of this section. In general, these conditions are necessary but not sufficient because of the inherent uncertainty in the definition of the monodromy charge which is given mod $1$.  Nevertheless, the beauty of this general argument  is that starting from an arbitrary \twozero{} model we get a handle on the massless spectrum in any twisted sector of any model in the family. 

\section{An example}\label{sec:example}
The fact that these conditions are necessary provides a prime test for where \emph{not} to look for massless states in a particular model. This can be of great importance when performing a computer scan in the space of models, so we give an example below.

Our starting point is the Gepner model $k^r=2^6$, which is a (2,2) model. In this model the internal CFT is a product of 6 minimal models each of which has central charge $c=\frac{3k}{k+2}=\frac{3}{2}$. All states will be of the form (\ref{eq:1}) but this time the internal CFT state is completely described by three vectors $\vec{l},\vec{q}$ and $\vec{s}$ so we will be using the notation $\Phi_\R=(w)(\vec{l},\vec{q},\vec{s})(p=0)$ instead. For the sake of the argument let us focus our attention on the massless charged spectrum in this model, which of course will fall into the fundamental ($\bf{27}$) or anti-fundamental  ($\bf{\overline{27}}$) representation of $E_6$. Without loss of generality, we will study states in the $\bf{27}$, which under the $SO(10)$ group decomposes into $\bf{10}+\bf{16}+\bf{1}$. Let us briefly remind the reader that the \righto-moving part of such massless states will then be of the form: 
\begin{itemize}
\item \textbf{10}s: $\Phi_\R=(v)(\Phi^I)(p=0)$ with 
\begin{equation*}\Phi^I\in\left\{  \underline{(0,0,0)^4 (0,2,2)^2},\ \underline{(0,0,0)^2 (1,-1,0)^4},\  \underline{(0,0,0)^3 (0,2,2)(1,-1,0)^2}    \right\}\ ,\end{equation*}

\item \textbf{16}s: $\Phi_\R=(c)(\Phi^{II})(p=0)$ with 
\begin{equation*}\Phi^{II}\in\left\{  \underline{(0,-1,-1)^4 (0,1,1)^2},\ \underline{(0,-1,-1)^2 (1,-2,-1)^4},\  \underline{(0,-1,-1)^3 (0,1,1)(1,-2,-1)^2}    \right\}\ ,\end{equation*}

\item \textbf{1}s: $\Phi_\R=(w=0)(\Phi^{III})(p=0)$ with 
\begin{equation*}\Phi^{III}\in\left\{  \underline{(0,-2,-2)^4 (0,0,0)^2},\ \underline{(0,-2,-2)^2 (1,-3,-2)^4},\  \underline{(0,-2,-2)^3 (0,0,0)(1,-3,-2)^2}    \right\}\ ,\end{equation*}
\end{itemize}
where underlining means permutations.

In this model $\beta_0$ has the usual form
\beq\label{eq:20}
\beta_0=(c)(0,1,1)^6(p=0)
\eeq
and is of order $N_{\beta_0}=8$. We choose the simple current with which we will orbifold our theory to be
\beq
\beta=\ (w=0)(2,1,-1)(0,0,0)^5(p=0)\ ,
\eeq
which is also of order $N_{\beta}=8$ and we note that $Q_{\beta}(\beta_0)=\frac{3}{8}\notin\Z{}$. Therefore the gauge bosons extending $SO(10)$ to $E_6$ are indeed projected out and we end up with a \twozero{} model. As explained in the previous section, this process naturally induces a whole family of models  ${\cal{M}}_0,\cdots,{\cal{M}}_7$ that arise if we orbifold by $\beta, \cdots,\beta+7\beta_0$ respectively.

The untwisted sector in all of these models will be the same and it will consist of all the states mentioned above that satisfy the invariance condition
\beq
Q_{\beta}(\Phi_\R)\in\Z{}\quad \Leftrightarrow\quad \frac{-2l_1+q_1+2s_1}{8}\in\Z{}.
\eeq 
For the $n-$twisted sector we will use equation (\ref{eq:condition2}). $h(\beta)$ can be readily calculated from the known formula for Gepner models\cite{Gepner}:
\beq
h=\sum_{i=1}^r\left(\frac{l_i(l_i+2)-q_i^2}{4(k_i+2)}+\frac{s_i^2}{8}\right)
\eeq
and we find that
\beq
n\Big(h(\beta)+\frac{1}{2}Q_{\beta}(\beta)\Big)=n\Big(\frac{9}{16}+\frac{1}{2}(-\frac{5}{8})\Big)=\frac{n}{4}\in\Z{}.
\eeq
This means that massless states can only arise in the untwisted $n=0$ sector, which we have already studied, or in the $n=4$ twisted sector. In the latter sector the \righto-moving part of the states will be of the form
\beqn
\Phi_\R&=&\tilde\Phi_\L+4(\beta+m\beta_0)\nonumber\\
&=&\tilde\Phi_\L+4\beta+4m\beta_0\nonumber\\
&=&\tilde\Phi_\L+(w=0)(0,4,0)(0,0,0)^5(p=0)+m(w=0)(0,4,0)^6(p=0)\nonumber\\
&=&\begin{cases} \Phi_\L+(w=0)(0,4,0)(0,0,0)^5(p=0) &\mbox{if m even}\label{eq:twistedexample}\\
 \Phi_\L+(w=0)(0,0,0)(0,4,0)^5(p=0) &\mbox{if m odd} \end{cases}\ , 
\eeqn
where we have used the properties\cite{Gepner} that for Gepner models $q$ is defined mod $2(k+2)$, $s$ is defined mod $4$ and we have also performed the identification $(l,q,s)\equiv(k-l,q+k+2,s+2)$ multiple times. A quick comparison with $\Phi^{I}$, $\Phi^{II}$ and $\Phi^{III}$ given above shows that states of the form (\ref{eq:twistedexample}) cannot be massless charged states, so the spectrum consists of the states in the untwisted sector only.

Once more, the power of this method is that it allowed us to check only one twisted sector ($n=4$) for massless states, as opposed to checking as many as seven of them for each model that we would \textit{a priori} expect in this example.

\section{Some further generalizations}\label{sec:generalizations}
There are many ways to generalize the above ideas to generate even more relationships in the space of \twozero{} models. For example, we are not restricted to using only $\beta_0$ but the natural splitting of the states into an $SO(10)$ part, an internal $N=2$ CFT and an $E_8$ part suggests that any
$$\beta_{0'}=(w)(\beta_0^{\text{CFT}})(p)$$
would generate its own orbit of \twozero{} models. Furthermore, when the internal CFT can be written as a tensor product of $N=2$ superconformal theories each term comes with a spectral flow operator $\beta_0^i$. We can then go one step further and use only some of the $\beta_{0}^i$'s instead of the entire $\beta_{0}^{\text{CFT}}$. 

Finally, as explained earlier, the presence of a simple current $J$ that breaks (2,2) to \twozero{} increases the possibilities even further. We can now have any linear combination of $J$, with any of the $\beta$'s mentioned above, with or without discrete torsion, and any such simple current will create its own orbit in the space of \twozero{} models.

In this paper we have shown explicitly how to use one of these mappings, the spectral flow $\beta_0$, to generate an entire family of models and we have derived useful expressions for the analysis of the spectra of these models. We believe that having not just one, but a big selection of such mappings as explained above will prove to be an important tool in the classification of \twozero{} models. 

\section{Conclusions}\label{sec:conclusions}
Heterotic-string vacua with \twozero{} world-sheet supersymmetry
are particularly interesting from a phenomenological 
point of view, as they reproduce the $SO(10)$ 
GUT structure, which is well motivated by the Standard Model
data. Ultimately, the confrontation of a string vacuum
with low scale experimental data will be achieved 
by associating it with an effective smooth quantum 
field theory limit. However, while the moduli spaces 
of (2,2) heterotic-string compactifications, and consequently their 
smooth limit, are reasonably well understood, this is not the case for those
with \twozero{} world-sheet supersymmetry. Indeed, the study of these
moduli spaces is an area of intense contemporary research \cite{twozeromoduli}. 

In this paper we discussed how the spinor-vector duality, which was 
observed in the framework of heterotic-string compactifications
with free world-sheet CFTs, can be extended to those with 
general RCFTs. The recipe adopted from the free case is the 
following: We start with a (2,2) compactification and break the world-sheet
supersymmetry on the bosonic side. The spectral 
flow operator, that operates as a symmetry generator of the 
(2,2) theory, then induces a map between the string vacua 
of the \twozero{} theory. As such, the map induced by the 
spectral flow operator provides a useful tool to explore 
the moduli spaces of \twozero{} heterotic-string compactifications. 
The question of interest in this respect is twofold. First, 
is this description complete? Namely, do all \twozero{} heterotic-string
compactifications descend from (2,2) theories? Second, 
what is the imprint of this map in the effective field 
theory limit? We hope to return to these questions in 
future publications. 

\section*{Acknowledgments}
A.E.F. thanks the Weizmann Institute and the Theoretical Physics
Department at the University of Oxford for hospitality.
P.A. acknowledges support from the Hellenic State Scholarships
Foundation (IKY). 
This work was supported in part by the STFC (ST/J000493/1).

\end{document}